\documentclass[final,5p,times,twocolumn]{elsarticle}
\usepackage{graphicx}
\usepackage{amsmath}
\usepackage{amssymb}
\usepackage{bbm,array}

\journal{Physica E}

\begin{document}

\newcommand{\be}{\begin{equation}}
\newcommand{\beq}{\begin{equation}}
\newcommand{\ee}{\end{equation}}
\newcommand{\bea}{\begin{eqnarray}}
\newcommand{\eea}{\end{eqnarray}}
\newcommand{\ba}{\begin{array}}
\newcommand{\ea}{\end{array}}
\newcommand{\Id}[1] {\int \! \! {\rm d}^3 #1}
\newcommand{\ID}[1] {\int \! \! \! \frac{{\rm d}^3 #1}{(2 \pi)^3}}
\newcommand{\citenew}[1] {\refnote{\cite{#1}}}
\newcommand{\w}{\omega}
\newcommand{\g}{\gamma}
\newcommand{\G}{\Gamma}
\newcommand{\vr} {{\bf r}}
\newcommand{\vj} {{\bf j}}
\newcommand{\vp} {{\bf p}}
\newcommand{\vs} {{\bf s}}
\newcommand{\nup}{n_{\uparrow}}
\newcommand{\ndown}{n_{\downarrow}}
\def\v#1{\mbox{\boldmath $#1$}}
\newcommand{\vnu} {{\bf \nu}}

\begin{frontmatter}

\title{Exchange and correlation energy functionals for two-dimensional open-shell systems}

\author[label1]{E.~R\"as\"anen}
\ead{erasanen@jyu.fi}
\author[label2,label3]{S. Pittalis}

\address[label1]{Nanoscience Center, Department of Physics, University of
  Jyv\"askyl\"a, FI-40014 Jyv\"askyl\"a, Finland}
\address[label2]{Institut f{\"u}r Theoretische Physik, Freie
  Universit{\"a}t Berlin, Arnimallee 14, D-14195 Berlin, Germany}
\address[label3]{European Theoretical Spectroscopy Facility (ETSF)}

\begin{abstract}
We consider density functionals for exchange and correlation energies
in two-dimensional systems. 
The functionals are constructed by 
making use of exact constraints for 
the angular averages of the corresponding exchange and
correlation holes, respectively, and assuming proportionality
between their characteristic sizes.
The electron current and spin are explicitly taken
into account, so that the resulting functionals
are suitable to deal with systems exhibiting orbital 
currents and/or spin polarization.  
Our numerical results show that in finite systems the
proposed functionals outperform the standard
two-dimensional local spin-density approximation, still performing 
well also in the important limit of the homogeneous two-dimensional 
electron gas.
\end{abstract}

\end{frontmatter}

\section{Introduction} \label{Intro}

Practical success of density-functional theory~\cite{dft} (DFT) 
depends on finding good approximations for exchange and 
correlation energy functionals. So far, most density
functionals have been developed in three dimensions (3D) with a view
to studying the properties of atoms, molecules, and solids. 
Apart from very recent progress~\cite{xh2D,ch2D,ring,cs,gga,simple,gamma},
such efforts for two-dimensional (2D) systems have been relatively
scarce beyond the commonly-used 2D local spin-density approximation (LSDA).
However, the rapidly increasing theoretical, experimental, and technological 
interest in 2D structures such as semiconductor layers and surfaces, 
quantum Hall systems, graphene, and various types of quantum dots,
calls for further developments.

Here we consider expressions of density functionals for
exchange and correlation energies of 2D systems.
These approximations are derived~\cite{xh2D,ch2D} by modeling
the angular average of the exchange and correlation
holes and using first-principle arguments
along the lines of Refs.~\cite{Becke_Roussel,Becke_other}.
The resulting spin- and current-dependent density functionals 
are then combined to obtain a new approximation for
the exchange plus correlation energy beyond the LSDA. 
Numerical tests, performed on systems for which
accurate reference data is available, show that
our functionals are superior to the 
LSDA. Moreover, we find a good agreement with the exact result
for the homogeneous 2D electron gas (2DEG).

\section{Basic formalism}

Within the Kohn-Sham (KS) method of spin-DFT,~\cite{BarthHedin:72} the ground 
state energy and spin densities 
$\rho_{\uparrow}(\vr)$ and $\rho_{\downarrow}(\vr)$ 
of a system of $N=N_{\uparrow}+N_{\downarrow}$ interacting 
electrons are determined. The formalism is general in terms of
dimensionality and the form of interaction. However, here we focus
on Coulomb-interacting particles in 2D.
The total energy, which is minimized to obtain the
ground-state energy, is written as a 
functional of the spin densities (in Hartree atomic units)
\begin{eqnarray}
E_{v}[\rho_{\uparrow},\rho_{\downarrow}] & = & T_s[\rho_{\uparrow},\rho_{\downarrow}] + 
E_{\rm H}[\rho] + E_{xc}[\rho_{\uparrow},\rho_{\downarrow}] \nonumber \\
& + & \sum_{\sigma=\uparrow,\downarrow}\int{d\vr} \; v_{\sigma}(\vr)
\rho_{\sigma}(\vr), 
\label{etot}
\end{eqnarray}
where $T_s[\rho_{\uparrow},\rho_{\downarrow}]$ 
is the KS kinetic energy functional, 
$v_{\sigma}(\vr)$ is the external (local) scalar potential acting 
upon the interacting system, $E_{\rm H}[\rho]$ 
is the classical electrostatic or Hartree energy of the total charge density 
$\rho(\vr)=\rho_{\uparrow}(\vr)+\rho_{\downarrow}(\vr)$, and  
$E_{xc}[\rho_{\uparrow},\rho_{\downarrow}]$ is the exchange-correlation 
energy functional. The latter can be further decomposed
into the exchange and correlation parts as
\be
E_{xc}[\rho_{\uparrow},\rho_{\downarrow}] = 
E_{x}[\rho_{\uparrow},\rho_{\downarrow}] + E_{c}[\rho_{\uparrow},\rho_{\downarrow}].
\ee
This is the central quantity which is formally well 
defined and exact, but needs to be approximated in practical 
applications of DFT.

\section{Modeling the exchange}\label{exchange}

The exchange-energy functional can be expressed as
\begin{equation}
E_x[\rho_{\uparrow},\rho_{\downarrow}] =  \frac{1}{2} \sum_{\sigma} \int d\vr \rho_{\sigma}(\vr) 
U^{\sigma}_x(\vr),
\label{xenergy}
\end{equation}
where
\begin{equation}
U^{\sigma}_x(\vr)=  2 \pi \int_{0}^{\infty}  ds \,\bar{h}^{\sigma}_x(\vr,s)
\label{xenergydensity}
\end{equation}
is the exchange-hole potential, and 
\begin{equation}
\bar{h}^\sigma_x(\vr,s) =  \frac{1}{2\pi}
\int_{0}^{2\pi} d\phi_s  h^\sigma_x(\vr,\vr+\vs)
\label{avxhole}
\end{equation}
is the 2D cylindrical average of the exchange-hole function
\begin{equation}
h^{\sigma}_{x}(\vr_1,\vr_2) = -
\frac{|\sum_{k=1}^{N_\sigma}\psi^*_{k,\sigma}(\vr_1)\psi_{k,\sigma}(\vr_2)|^2}
{\rho_{\sigma}(\vr_1)},
\label{xhole}
\end{equation}
where the sum in the numerator is the
one-body spin-density matrix of the (ground state) Slater 
determinant constructed from
the KS orbitals, $\psi_{k,\sigma}$. 
As the basis of our exchange model~\cite{xh2D}, we
consider
\begin{equation}
\bar{h}^{\sigma}_x(a,b;s) = - \frac{a}{\pi} \exp \left[-a \left( b + s^2 \right) \right] I_0(2a\sqrt{b}s)\;,
\label{h}
\end{equation}
where $I_0(x)$ is the zeroth order modified Bessel function of the first kind
(note that $I_0(0)=1$).
This function provides a non-positive  
exchange hole and satisfies the normalization
\begin{equation}
\int d\vs h^{\sigma}_x(\vr,\vs) = -1.
\label{norm}
\end{equation}
In Eq. (\ref{h}), $a$ and $b$ are position-dependent functions, 
which are chosen 
to locally reproduce the short-range behavior of the exact exchange
hole. Comparison to the Taylor expansion leads to relations 
$a = \pi\rho_{\sigma}e^{y}$ and $b = y/a$,
where $y:=ab$ is the solution of 
\begin{equation}
\left(y-1\right)\exp(y) = \frac{C^{\sigma}_{x}}{\pi\rho^2_{\sigma}},
\label{relation}
\end{equation}
where
\begin{equation}\label{C}
C^{\sigma}_x = \frac{1}{4}\left[ \nabla^2 \rho_{\sigma} -2\tau_\sigma
+ \frac{1}{2}\frac{\left( \nabla \rho_\sigma \right)^2}{\rho_\sigma}
+ 2 \frac{\vj^2_{p,\sigma}}{\rho_\sigma} \right]
\end{equation}
is the so-called curvature of the exchange hole, $\tau_\sigma$
is (twice) the spin-dependent kinetic-energy density, and $\vj_{p,\sigma}$
is the spin-dependent paramagnetic current density.
Both $\tau_\sigma$ and
$\vj_{p,\sigma}$
depend explicitly on the KS orbitals. Thus the expression in Eq.~(\ref{C})
has an implicit dependence on the spin-densities $\rho_\sigma$.

Note that the model introduced in Eq.~(\ref{h}) is exact 
in the case of the ground-state single-electron 
wave function of a 2D harmonic oscillator.
Furthermore, application of the above model to the 2DEG 
yields the exchange energy per particle as
\begin{equation}
\epsilon_{x}[r_s,\xi]=-\frac{\sqrt{\pi}}{4\sqrt{2}\,r_s}\left[(1+\xi)^{3/2}+(1-\xi)^{3/2}\right],
\label{final_xenergy_N}
\end{equation}
where $r_s=1/\sqrt{\pi\rho}$ and
$\xi=(\rho_{\uparrow}-\rho_{\downarrow})/\rho$ is the spin polarization.
This energy
is about $4.4\%$ smaller than the exact one~\cite{rajagopal}. 
Interestingly, Eq.~(\ref{final_xenergy_N}) may be seen as a 
correction of the LSDA exchange functional suitable for dealing with 
finite few-electron systems.

\section{Modeling the correlation}\label{correlation}

The correlation-energy functional can be expressed as follows 
\be
E_c [\rho_{\uparrow},\rho_{\downarrow}] =E^{\uparrow\uparrow}_{c} [\rho_{\uparrow},\rho_{\downarrow}] + E^{\downarrow\downarrow}_{c} [\rho_{\uparrow},\rho_{\downarrow}] + 
2 E^{\uparrow\downarrow}_{c} [\rho_{\uparrow},\rho_{\downarrow}],
\ee
where we have used $E^{\uparrow\downarrow}_{c} [\rho_{\uparrow},\rho_{\downarrow}]=E^{\downarrow\uparrow}_{c} [\rho_{\uparrow},\rho_{\downarrow}]$
and
\be
E^{\sigma\sigma'}_{c}[\rho_{\uparrow},\rho_{\downarrow}] = \pi  \int d\vr\, \rho_{\sigma}(\vr)  \int_{0}^{\infty} ds \, \bar{h}^{\sigma\sigma'}_{c}(\vr,s)
\ee
with $\bar{h}^{\sigma\sigma'}_{c}(\vr,s)$ being  the cylindrical average of the correlation-hole function,
$ h^{\sigma\sigma'}_{c}(\vr_1,\vr_2)$, computed in the same way as  $\bar{h}^\sigma_x(\vr,s)$.
We remind that the correlation-hole functions can be obtained by the coupling constant integration
\be
h^{\sigma\sigma'}_{c}(\vr_1,\vr_2) = 
\int_{0}^{1} d\lambda \, h^{\sigma\sigma'}_{c,\lambda}(\vr_1,\vr_2), 
\ee
where $ h^{\sigma\sigma'}_{c,\lambda}(\vr_1,\vr_2) = h^{\sigma\sigma'}_{\lambda}(\vr_1,\vr_2) - h^{\sigma}_{x}(\vr_1,\vr_2)\delta_{\sigma\sigma'}$,
$\{\sigma\sigma'\}=\{\uparrow\uparrow,\downarrow\downarrow,\uparrow\downarrow,\downarrow\uparrow\}$,
and the parameter $\lambda\in[0,1]$ is the electronic coupling
strength~\cite{dft}.
Now, we model ~\cite{ch2D} the cylindrical averages of the $\lambda$-dependent 
correlation-hole functions as
\be\label{mch1}
\bar{h}^{\sigma\sigma}_{c,\lambda}(\vr,s)  = 
\frac{2}{3} \lambda D_{\sigma}(\vr) 
\left[ \frac{s-z_{\sigma\sigma}(\vr)}{1+ \frac{2}{3}  \lambda z_{\sigma\sigma}(\vr) } \right]
s^2\,F\left(\gamma_{\sigma\sigma}(\vr)\,s\right),
\ee
and
\be\label{mch2}
\bar{h}^{\sigma{\bar \sigma}}_{c,\lambda}(\vr,s)  = 
2 \lambda \rho_{{\bar \sigma}}(\vr)
\left[ \frac{s-z_{\sigma{\bar \sigma}}(\vr)}{1+ 2  \lambda z_{\sigma{\bar \sigma}}(\vr) } \right]
F(\gamma_{\sigma{\bar \sigma}}(\vr)s) \;.
\ee
for the same spin and opposite spin case, respectively 
(note that, $\bar{\sigma}$ indicates
an opposite value w.r.t. $\sigma$). In Eq. (\ref{mch1}), 
$D_{\sigma}:= 1/8 \left( \nabla^2 \rho_\sigma - 4 C^{\sigma}_x
\right)$. This quantity vanishes for all the single-particle
($N=1$) systems, and thus also 
the corresponding (same-spin) correlation hole and correlation energy
vanish as in the exact case.
For $s \rightarrow 0$,  Eqs.~(\ref{mch1}) and (\ref{mch2}) recover the 
exact short-range behavior of the $\lambda$-dependent correlation-hole functions. $F$ is introduced 
to ensure  the decay of the correlation holes in the limit $s
\rightarrow \infty$. In the case of finite 2D systems,
it seems appropriate to set $F(x):=\exp(-x^2)$.
Furthermore, the exact condition \be\label{sumrule}
\int ds \,s\,\bar{h}^{\sigma\sigma'}_{c,\lambda}(\vr,s) = 0,
\ee
is satisfied by setting $\gamma_{\sigma\sigma} = 3\sqrt{\pi}/(4
z_{\sigma\sigma})$ and
$\gamma_{\sigma{\bar \sigma}} = \sqrt{\pi}/(2 z_{\sigma{\bar
    \sigma}})$.
Functions $z_{\sigma\sigma}$ and $z_{\sigma\bar{\sigma}}$ are defined as
the characteristic sizes of the correlation holes,  for which 
$\bar{h}^{\sigma\sigma}_{c,\lambda}(\vr,z_{\sigma\sigma})=0$, $
\bar{h}^{\sigma\bar{\sigma}}_{c,\lambda}(\vr,z_{\sigma\bar{\sigma}})=0$.
As for the 3D case,~\cite{Becke_other} we set 
$z_{\sigma\sigma} := 2 c_{\sigma\sigma} |U_x^{\sigma}|^{-1}$ and
$z_{\sigma{\bar \sigma}}:= c_{\sigma{\bar \sigma}} \left[ |U_x^{\sigma}|^{-1}+ |U_x^{{\bar \sigma}}|^{-1}\right]$
where $U_x^{\sigma}$ is the exchange-hole potential calculated in the
previous section, and 
$c_{\sigma\sigma}$ and $c_{\sigma{\bar \sigma}}$ are constants
to be determined (see below). In this way,
the characteristic size of the correlation hole is
proportional to the characteristic size of the corresponding exchange hole.
This is suggested by a picture in which electrons 
are correlated as long as their interaction is not screened
by their respective exchange holes. A more elaborate model
may require to consider $c_{\sigma\sigma'}$ as a density functional,
which is, however, beyond the scope of this work.

The approximations given above are {\em implicit} density functionals 
due to the {\em explicit} dependence on the KS orbitals
appearing in the expressions for $\tau$ and $\vj_{p,\sigma}$. Therefore,
these functionals can be
classified as current-dependent meta-generalized-gradient
approximations, and we expect them 
to be ideally suited to deal with a large class of spin-polarized 
and/or current-carrying ground-states of two-dimensional systems.

\section{Applications}

We test the present 2D functionals for a set
of $N$-electron parabolic quantum dots described by a Hamiltonian,
\begin{equation}                                                                
H = \sum^N_{i=1}\left(-\frac{1}{2}\nabla_i^2+\frac{1}{2}\omega^2 r_i^2\right)+\sum_{i<j}\frac{1}{|\vr_i-\vr_j|},
\label{hamiltonian}                                                             
\end{equation}
where $\omega=1/4$ is the confinement strength. 
The exchange energies obtained using the approximation given in
Sec.~\ref{exchange} are compared with the exact-exchange
(EXX) results within the Krieger-Li-Iafrate approximation~\cite{kli}.
Then, the correlation energies obtained according to
Sec.~\ref{correlation} are compared to the reference results
$E_c^{\rm ref}=E_{\rm tot}^{\rm ref}-E_{\rm tot}^{\rm EXX}$,
where $E_{\rm tot}^{\rm ref}$ is the numerically accurate
total energy of a configuration-interaction calculation
reported in Ref.~\cite{rontani}. The KS orbitals from  
self-consistent EXX calculations -- performed using the 
{\tt octopus} DFT code~\cite{octopus} -- are
used as the input for the functionals.

In Table~\ref{table} 
\begin{table*}
  \caption{\label{table} Exchange and correlation energies
for fully spin-polarized quantum dots. The last row shows
the mean percentage error $\Delta$ of our functional and the LSDA
for both exchange and correlation, respectively.
}
  \begin{tabular}{c c c c c c c c c}
  \hline
  \hline
$N$ & $L$ & $-E^{\rm EXX}_x$ & 
  $-E_{\rm x}^{\rm model}$ & $-E_x^{\rm LSDA}$ & $-E_c^{\rm ref}$ & $-E_c^{\rm model}$ & $-E_c^{\rm LSDA}$ \\
  \hline
  2  & 1 & 0.626  & 0.631  & 0.583 & 0.010 & 0.0108 & 0.0346 \\
  3  & 0 & 1.021  & 1.038  & 0.963 & 0.023 & 0.0212 & 0.0540 \\  
  4  & 2 & 1.374  & 1.428  & 1.332 & 0.034 & 0.0375 & 0.0731 \\
  5  & 0 & 1.816  & 1.864  & 1.745 & 0.048 & 0.0517 & 0.0931 \\
  6  & 0 & 2.213  & 2.267  & 2.126 & 0.064 & 0.0636 & 0.1125 \\
\hline
 $\Delta$ & &    & 2.3\,\% & 4.7\,\% &    & 6.9\,\% & 133\,\% \\
 \hline
\hline
  \end{tabular}
  \end{table*}
we show the results for fully spin-polarized quantum
dots with different $N=2\ldots 6$, spin $S=N/2$, and 
total angular momentum ({\em z} component) $L$. Note
that different values of $L$ correspond to different
paramagnetic current densities.
Overall, we find excellent performance in both the exchange
and correlation functionals. A significant improvement
over the LSDA as seen in the mean errors $\Delta$ on the last
row of Table~\ref{table}. It should be noted that the error
of the LSDA correlation is huge. This error is typically 
partially compensated by an error of an opposite sign in the LSDA exchange.
However, our functionals are able to reproduce {\em both}
the exchange and correlation energies with a good accuracy.

In the correlation functional we have used
a fixed value $c_{\sigma\sigma}=1.25$ as the prefactor in the 
characteristic size of the correlation hole 
(see the end of Sec.~\ref{correlation}). We point out that 
although this value is 
fundamentally a {\em post-hoc} choice, the 
variations in optimal $c_{\sigma\sigma}$ (and $c_{\sigma{\bar \sigma}}$)
for different systems are typically rather small~\cite{ch2D}.
This becomes obvious also within the 2DEG below, where 
results for various densities, corresponding to 
amounts of correlation, are shown with the
same fixed $c_{\sigma\sigma}$.

Finally we turn our attention from finite systems to the 2DEG.
Figure~\ref{gas}
\begin{figure}
\begin{center}\leavevmode
\includegraphics[width=0.99\columnwidth]{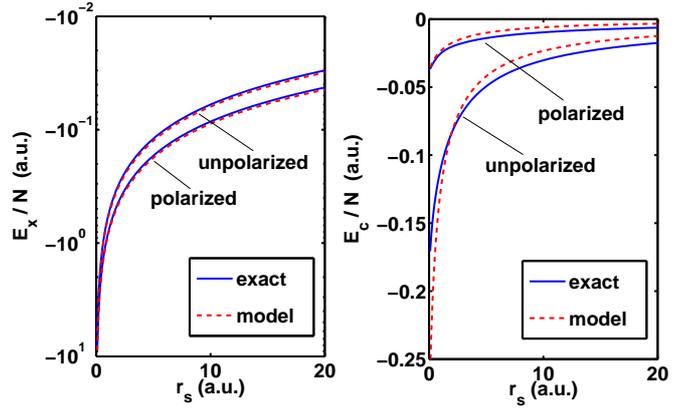}
\caption{(color online) Exchange (left) and correlation energies (right)
per particle as a function of the density parameter $r_s=1/\sqrt{\pi n}$
in the fully spin-polarized and spin-unpolarized homogeneous 
two-dimensional electron gas. The dashed lines show the results
 calculated with the present functionals. 
The solid lines correspond to the exact results obtained
from analytic calculations for the exchange~\cite{rajagopal} and
analytic parametrizations of the 
quantum Monte Carlo data for the correlation~\cite{attaccalite}.
Note the logarithmic scale in the exchange energies.
}
\label{gas}
\end{center}
\end{figure}
shows the exchange (left) and correlation energies (right) per
particle in the 2DEG as a function of the density 
parameter $r_s=1/\sqrt{\pi n}$ 
against exact results available from analytic~\cite{rajagopal} 
and quantum Monte Carlo calculations~\cite{attaccalite}, respectively.
As pointed out in Sec.~\ref{exchange}, in this limit our exchange functional
deviates consistently $4.4\%$ from the exact (LSDA) result.
Also for the correlation the agreement is very good up
to strong correlations -- it should be noted that
in quantum dots, for example, typical densities 
correspond roughly to $1\lesssim r_s \lesssim 5$.
In the correlation functional we have consistently used fixed values 
$c_{\sigma\sigma}=1.25$, i.e., the same value as with the quantum 
dots above, and $c_{\sigma{\bar \sigma}}=0.75$.

\section{Conclusions}

We have presented two-dimensional 
density functionals for 
exchange and correlation energies.
The functionals have been
derived by modeling the angular averages of the
corresponding exchange and correlation holes along
first-principle arguments and by enforcing known
exact constraints.
Tests of our functionals for
few-electron open-shell quantum dots and the homogeneous
two-dimensional electron gas show very good performance.
Hence, we may propose an accurate alternative to
the commonly used local spin-density approximation
for electronic-structure calculations in two dimensions.

This work was supported by the Academy of Finland,
Magnus Ehrnrooth Foundation,
Deutsche Forschungsgemeinschaft, and 
the EU's Sixth Framework
Programme through the ETSF e-I3.

\end{document}